# Destruction of adiabatic invariance at resonances in slow-fast Hamiltonian systems

A.I.Neishtadt and A.A.Vasiliev

*Space Research Institute, Moscow*

*Abstract*
There are many problems that lead to analysis of dynamical systems in which one can distinguish motions of two types: slow one and fast one. An averaging over fast motion is used for approximate description of the slow motion. First integrals of the averaged system are approximate first integrals of the exact system, i.e. adiabatic invariants. Resonant phenomena in fast motion (capture into resonance, scattering on resonance) lead to inapplicability of averaging, destruction of adiabatic invariance, dynamical chaos and transport in large domains in the phase space. In the paper perturbation theory methods for description of these phenomena are outlined. We also consider as an example the problem of surfatron acceleration of a relativistic charged particle.

## INTRODUCTION

Many problems in physics can be reduced to analysis of systems possessing motions on two time scales, namely fast and slow (slow-fast systems). An approximate integral of such a system is called its adiabatic invariant. It changes only a little on large time intervals, such that variation of slow variables is not small. One of well known examples is a pendumulum with slowly varying length and small amplitude. On long enough time intervals its length $l$ and the amplitude of its oscillations $A$ change considerably, but their combination $Al^{3/4}$ stays approximately constant. Another classical example is motion of a charged particle in slowly inhomogeneous magnetic field, i.e. a field varying a little on distances of order of the Larmor radius. In this motion the ratio of squared orthogonal velocity component to the field strength is approximately constant along the particle's trajectory. This property is used in magnetic traps.

Adiabatic invariants are important dynamical values. In particular, if a system has enough number of adiabatic invariants then the motion over long time intervals is close to regular. On the other hand, destruction of adiabatic invariance is one of mechanisms for onset of chaotic dynamics.

Adiabatic invariants usually arise as integrals of averaged systems. A slow-fast system can often be written as a system with rotating phases:

$$\dot{x} = \varepsilon f(x, \varphi, \varepsilon),$$
$$\dot{\varphi} = \omega(x) + \varepsilon g(x, \varphi, \varepsilon), \quad (1)$$

where $x, \varphi, \omega, f$, and $g$ are vectors, and $\varepsilon$ is a positive small parameter characterizing the rate of variation of slow variables $x$. Variables $\varphi$ are fast phases. They rotate with frequencies $\omega$, and functions $f$ and $g$ depend on them $2\pi$-periodically. For approximate description of motion in (1), one can use the averaging method, see, e.g. [1, 2], i.e. just to use instead of the first equation in (1) averaged equation

$$\dot{x} = \varepsilon F(x), \qquad F = \langle f \rangle^{\varphi}_{\varepsilon=0}. \quad (2)$$

Here $F$ is average of $f$ over $\varphi$ at $\varepsilon=0$. If the averaged system approximately describes the motion in the exact system and has a first integral $I(x)$, this first integral is an approximate integral of system (1), i.e. its adiabatic invariant. However, this is valid provided that the averaging method works. This is the case, for example, in one-frequency systems where the phase is scalar, like in the examples mentioned above [1, 2]. We are going to discuss multi-frequency systems. In this case, one encounters problems with resonances.

A resonant condition $k_1\omega_1(x) + ... + k_n\omega_n(x) = 0$, where $k = (k_1,...,k_n)$ is an integer and non-zero vector and $\omega = (\omega_1,..., \omega_n)$, defines a surface in the space of slow variables $x$, called the resonant surface. Frequencies $\omega$ depend on $x$, and as the latter slowly change, a phase point of the averaged system can cross the resonant surface. Behaviour of the corresponding phase point in the exact system can be quite different. We put forward the theory of resonant phenomena for Hamiltonian two-frequency slow-fast systems and consider a sample problem of surfatron acceleration of a relativistic charged particle.

## SLOW-FAST HAMILTONIAN SYSTEMS

Consider a Hamiltonian system with Hamiltonian

$$H = H_0(p, q, I) + \varepsilon H_1(p, q, I, \varphi, \varepsilon), \quad (3)$$

where $\varepsilon$ is a positive small parameter and canonically conjugated pairs of variables are

$$(p, \varepsilon^{-1}q) \in \mathbb{R}^{2n}, \quad \text{and } (I, \varphi) \in \mathbb{R}^m \times \mathrm{T}^m.$$

The corresponding equations of motion are:

$$\dot{p} = -\varepsilon \frac{\partial H_0}{\partial q} - \varepsilon^2 \frac{\partial H_1}{\partial q}$$

$$\dot{q} = \varepsilon \frac{\partial H_0}{\partial p} + \varepsilon^2 \frac{\partial H_1}{\partial p}$$

$$\dot{I} = -\varepsilon \frac{\partial H_1}{\partial \varphi} \qquad (4)$$

$$\dot{\varphi} = \frac{\partial H_0}{\partial I} + \varepsilon \frac{\partial H_1}{\partial I}$$

The first three equations describe evolution of slow variables, and the last equation describe behaviour of fast phases. Systems like (4) often arise as follows. Suppose a Hamiltonian system has fast and slow variables, and at a frozen value of the slow variables the remaining system for the fast variables (called the fast system) is integrable. Hence, standard "action-angle" variables can be introduced in the fast system. In this case, the Hamiltonian of the complete system can be reduced to the form (3), see, e.g. [3, 4]. To average system (4) one averages Hamiltonian (3). The averaged Hamiltonian does not depend on $\varphi$, and the averaged system (called in this case the system of adiabatic approximation) looks as follows:

$$\dot{p} = -\varepsilon \frac{\partial H_0}{\partial q}, \quad \dot{q} = \varepsilon \frac{\partial H_0}{\partial p}, \quad \dot{I} = 0. \quad (5)$$

Hence, components of vector $I$ are integrals of the averaged system and good candidates to be adiabatic invariants of the exact system. Behaviour of $(p,q)$ in (5) is described by a Hamiltonian system depending on $I$ as a parameter, which can be completely studied in the important case $n=1$. Therefore, a phase trajectory of the adiabatic approximation belongs to a plane $I$ = const and is, in case $n=1$, an intersection of a level surface of $H_0$ with this plane. Suppose that this adiabatic trajectory crosses a resonant surface. How does the corresponding phase trajectory of the exact system behave at this crossing? It turns out that there can occur phenomena of two kinds.

The first one is capture into the resonance. The exact trajectory follows closely the adiabatic trajectory till the resonant surface, where the exact trajectory starts following this surface rather than the adiabatic trajectory. Thus, adiabatic invariant $I$ changes strongly along the exact motion. After a while, the trajectory may leave the resonant surface (escape from the resonance) and start following a different adiabatic trajectory with value of $I$ completely different from the original one. It turns out that initial conditions of trajectories to be captured and those to cross the resonance without capture are intermixed. Therefore, one should speak about capture probability. This probability is small, of order of $\varepsilon^{1/2}$, and only a small part of trajectories will be captured. However, if phase trajectories of the averaged system are closed (which is often the case), exact trajectories cross the resonant surface again and again, making the probability to be captured on a long time interval quite high. Hence this phenomenon is important for the dynamics. Capture into resonance was considered first in [5, 6] (for systems with dissipation).

Another phenomenon is scattering on the resonance. It takes place for trajectories that cross the resonant surface without capture. The exact trajectory follows closely the adiabatic trajectory, and at the resonant surface shifts a little bit (by a value of order of $\varepsilon^{1/2}$). After crossing the surface, the exact trajectory follows a different adiabatic trajectory, which is at a distance of order of $\varepsilon^{1/2}$ from the original one. For a different initial condition this variation in $I$ is different. Scattering on resonance was considered first in [7, 8]. Scattering on a resonance is also small, but again, if the trajectory crosses the resonance surface repeatedly, multiple scatterings result in diffusion of the adiabatic invariant. Thus, these two resonant phenomena determine dynamics on long time intervals.

We consider these phenomena in two-frequency systems, when $\varphi=(\varphi_1, \varphi_2)$, $\omega=(\omega_1, \omega_2)$, $I=(I_1, I_2)$. In such systems, resonant surfaces are arranged in a simple way: on the plane $(\omega_1, \omega_2)$, they are represented as straight lines with rational slopes passing through the origin. An adiabatic phase trajectory crosses them one by one, and the effect of each resonance can be studied separately.

## RESONANT PHENOMENA IN TWO-FREQUENCY SYSTEMS

Consider a resonant surface defined by $k_1\omega_1 + k_2\omega_2 = 0$, where $k_1, k_2$ are co-prime integer numbers. We describe Hamiltonian version of the standard reduction (see, e.g. [9]) of the system near the resonant surface. The first step is a standard one called partial averaging near the resonance. We introduce resonant phase $\gamma$, which varies slowly near the resonance and fast far from it. In a Hamiltonian system we do this with a canonical change of variables $(p, q; I_1, I_2, \varphi_1, \varphi_2) \rightarrow (p, q; R, J, \gamma, \chi)$, where

$$\gamma = k_1\varphi_1 + k_2\varphi_2, \quad R = l_2 I_1 - l_1 I_2,$$
$$\chi = l_1\varphi_1 + l_2\varphi_2, \quad J = -k_2 I_1 + k_1 I_2, \quad (6)$$

and $l_1, l_2$ are integers, $k_1 l_2 - k_2 l_1 = 1$. Now one can average over fast variable $\chi$ to obtain Hamiltonian

$$H = H_0(R, J, p, q) + \varepsilon \langle H_1 \rangle^\chi. \quad (7)$$

Here $J$=const and can be considered as a parameter. As we are interested in phenomena occuring near the resonant surface, we expand the Hamitonian in its small neighbourhood (of order of $\varepsilon^{1/2}$). The resonant surface is defined by $R=R_{\text{res}}(p, q)$. Introduce rescaled distance from this surface $P$ with a canonical transformation of variables $(p, q; R, \gamma) \rightarrow (\bar{p}, \bar{q}; P, \bar{\gamma})$, where

$$P = \frac{R - R_{\text{res}}(p,q)}{\sqrt{\varepsilon}} + O(\sqrt{\varepsilon}),$$
$$\bar{p} = p + O(\varepsilon), \quad \bar{q} = q + O(\varepsilon). \quad (8)$$

Introduce also rescaled time $\tau = \varepsilon^{1/2} t$ and denote the derivative over $\tau$ with a prime. After expanding the Hamiltonian one obtains a system of universal form [3], see also [4]. The equations describing motion along the resonant surface are decoupled, and for this motion we obtain a Hamiltonian system with Hamiltonian function $\Lambda$, which is the restriction of Hamiltonian $H_0$ onto the resonant surface (bars are omitted):

$$p' = -\sqrt{\varepsilon} \frac{\partial \Lambda}{\partial q}, \quad q' = \sqrt{\varepsilon} \frac{\partial \Lambda}{\partial p}. \quad (9)$$

We call phase flow of this system the resonant flow. For resonant phase $\gamma$ and conjugated variable $P$ one

obtains a 1 d.o.f. Hamiltonian system with a standard pendulum-like Hamiltonian function

$$F = \frac{1}{2}gP^2 + f + b\gamma, \qquad (10)$$

where

$$g = \left.\frac{\partial^2 H_0}{\partial P^2}\right|_{res}, \quad f = \left.\langle H_1 \rangle^\chi\right|_{res}, \quad b = \{R_{res}, \Lambda\},$$

with $\{\cdot,\cdot\}$ denoting a Poisson bracket w.r.t. $(p,q)$. Corresponding equation of motion is similar to one of a pendulum under the action of a rotatory torque:

$$\left(\gamma'/g\right)' = -\frac{\partial f}{\partial \gamma} - b. \qquad (11)$$

One can see that the first term in the right hand side is a periodic function of $\gamma$ with zero average (in case of a pendulum it is sine), and independent of $\gamma$ term $b$ is a rotatory torque. The parameters of this pendulum depend on slow variables $(p, q)$, which vary according to equations (9). As they vary slowly (at a speed of order $\varepsilon^{1/2}$), it is reasonable to begin with studying the pendulum-like system at frozen values of its parameters. There are two kinds of phase portraits of this system at different parameters values. The first one is when the torque is strong enough and the pendulum can only pass from rotation in one direction to rotation in the other direction (Fig. 1 a). The other case is when the torque is not that strong, and there are also separatrices and ocsillatory domains on the phase portrait (Fig. 1 b). The resonant surface on these pictures is represented by the axis $P=0$. The resonant phenomena of capture into the resonance and escape from the resonance, however, cannot be seen in these pictures. They occur due to slow variation of the pendulum's parameters. Because of these variations a phase trajectory close to a separatrix in Fig. 1 b can cross the separatrix and get inside the separatrix loop. This is a capture into the resonance. Otherwise, a trajectory inside the loop can cross the separatrix and escape. This is escape from the resonance. Consider these phenomena in more details.

The area inside the separatrix loop $S(p, q)$ on the phase portrait in Fig. 1 b is a function of a point $(p, q)$ on the resonant surface, which can be calculated. If it grows along trajectories of the resonant flow (9), there appears additional space inside the loop. As Hamiltonian systems conserve phase volume, additional phase points can enter the domain inside the separatrix loop. This means a possibility of the capture. Hence, the capture is possible at those points of the resonant surface, where function $S(p, q)$ grows along the resonant flow. On the contrary, at the points where this function decreases along the resonant flow, only escape from the resonance is possible.

There exists a rigorous definition of capture probability [3, 4], but here we give only an explanation. Take an initial point $M_0$ far from the resonant surface and consider a small ball centred at this point. The ratio of phase volume inside this ball that is captured into the resonance to the ball's total volume is the capture probability at $M_0$. This probability can be calculated according to the following formula ([3], see also [4]):

$$\Pr(M_0) = \begin{cases} \dfrac{\sqrt{\varepsilon}\{S,\Lambda\}_*}{2\pi |b_*|}, & \text{if } \{S,\Lambda\}_* > 0 \\ 0, & \text{if } \{S,\Lambda\}_* < 0. \end{cases} \qquad (12)$$

Here the asterisk denotes that corresponding values should be calculated at the point where the adiabatic trajectory starting at $M_0$ hits the resonant surface.

Once a phase point is captured into the resonance, it is interesting to know, where it will escape from the resonance, if ever. To construct this "in-out" function, we note that a captured phase point on the phase plane $(\gamma, P)$ performs fast rotation inside the oscillatory domain of a pendulum with slowly varying parameters (Fig.1 b). Again we have a Hamiltonian system with slowly varying parameters. This system has an adiabatic invariant (we call it the inner adiabatic invariant). This is the area surrounded by the phase trajectory at frozen values of the parameters. As time goes, the parameters change, but this area stays approximately the same. Consider, for simplicity, the case $n=1$; so, system for resonant flow (9) has 1 d.o.f. On the resonant surface we can draw two sets of level lines: those of function $S$ and those of function $\Lambda$. The latter are trajectories of the resonant flow. At a point where a phase trajectory hits the resonant surface, two level lines of these two sets intersect, corresponding to $S_*$ and $\Lambda_*$. If the point is captured, it continues its motion along the level line $\Lambda = \Lambda_*$, and function $S$ in this motion grows (at least for a while). On the other hand, the area surrounded by the phase trajectory on plane $(\gamma, P)$ stays approximately equal to $S_*$. Therefore, escape from the resonance occurs when the level line $\Lambda = \Lambda_*$ crosses the level line $S = S_*$ once again. This gives a simple receipt for construction of the "in-out" function.

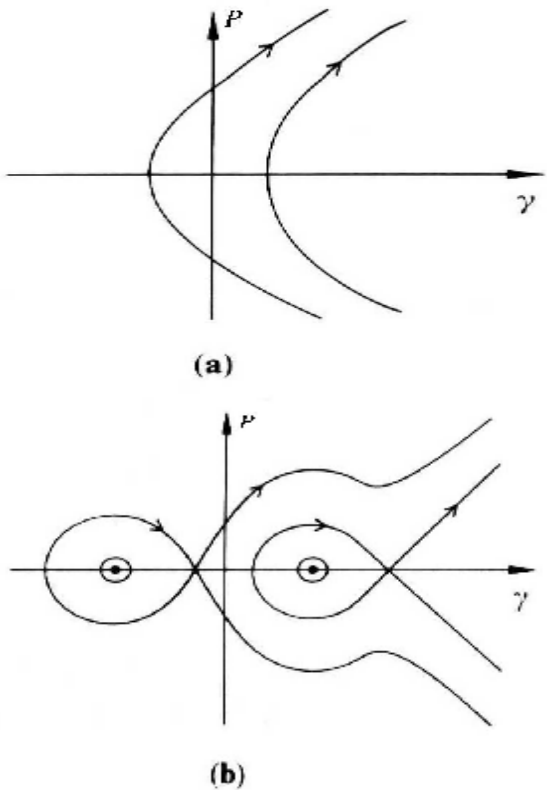

**FIGURE 1.** Phase portraits of the pendulum-like system.

To describe scattering on the resonance, one can consider equation of motion for variable $R$ (see (6),(7))

$$\dot R = -\varepsilon \frac{\partial \langle H_1 \rangle^\chi}{\partial \gamma}, \qquad (13)$$

and integrate it along the phase trajectory to find total variation of $R$. Far from the resonance under the integral we have a fast oscillating function, and the main contribution is given by a part of trajectory near the resonance. To calculate this contribution one can fix in the r. h. s. of (13) values of slow variables equal to that at which the adiabatic trajectory hits the resonant surface and integrate the obtained expression along the trajectory of pendulum (11) for these values of slow variables. Thus one obtains the following formula for $\Delta R$, see, e.g. [4]:

$$\Delta R = -2s\sqrt{\varepsilon} \int_{-s\infty}^{\gamma_*} \frac{\partial f/\partial \gamma}{\sqrt{2g(h_* - f - b\gamma)}} d\gamma + ..., \quad (14)$$

where $s=\text{sign}(bg)$. Here, $\gamma_*$ and $h_*$ are values of $\gamma$ and $F$ at which the exact trajectory hits the resonant surface. Dots denote high order terms. Value of $\Delta R$ depends on value of the resonant phase at the time when the trajectory hits the resonant surface. This phase is difficult to predict, because far from the resonant surface it rotates fast, and makes many turns before the point gets to the resonant surface. Therefore, in a sense, this value should be considered as random one. Instead of this phase, it is convenient to use value $\xi$, defined as

$$\xi = \text{Frac}(h_*/2\pi|b_*|). \qquad (15)$$

Here Frac denotes the fractional part. This value can be treated as random, because small change of initial conditions produces large variation of $\xi$. Then the probability for $\xi$ to be between certain numbers $a$ and $b$, $0 < a < b < 1$, can be defined as a relative measure of initial conditions such that $\xi$ gets into this interval. It turns out that this probability is equal to $b - a$, i.e. $\xi$ is uniformly distributed on the interval $(0,1)$. Hence, formula (14) gives distribution of variation of adiabatic invariant due to passage through the resonance.

Consider now consecutive crossings of different resonances or consecutive multiple crossings of the same resonance. On each crossing the phase trajectory undergoes scattering on the resonance and the value of the adiabatic invariant changes. Take two consecutive crossings. If we change the initial value of $\gamma$ by a small $\Delta\gamma$, the resulting change in the variation of the adiabatic invariant is of order $\varepsilon^{1/2} \Delta\gamma$. The system is nonlinear, and frequency of motion depends on the slow variables. Therefore this change of adiabatic invariant produces variation of frequency by a value of the same order. It takes time of order $1/\varepsilon$ to hit the resonant surface next time. In this time period, the resonant phase changes by a value of order $(1/\varepsilon)\varepsilon^{1/2}\Delta\gamma = \varepsilon^{-1/2} \Delta\gamma$, which is much larger than initial variation $\Delta\gamma$. According to a heuristic but generally accepted criterion of phase stretching, we can consider values of the resonant phase at successive resonance crossings as statistically independent. Together with formula (14) giving the distribution function for variation of the adiabatic invariant on a single crossing, formula (12) for probability of capture into resonance and description of the "in-out" function, this gives a complete statistical description of motion. It looks like a sort of random walk, and results in diffusion of the adiabatic invariant on long time intervals. For some cases proof of independence of results of passages through a resonance on time interval of the length $1/\varepsilon^{3/2}$ is given in [10].

## AN EXAMPLE: SURFATRON ACCELERATION

In this section we illustrate the theory presented above with an example of motion of a relativistic charged particle in stationary uniform magnetic field $\boldsymbol{B}$ and an oblique high frequency harmonic electrostatic wave with amplitude $E_0$ and potential $\Phi$. In [11], possibility of so called unlimited surfatron acceleration in this system was shown: under certain conditions a particle initially put into the potential well of the wave stays in this well and moves along the wave front like a surfer with acceleration, and can be accelerated up to the speed of light. This problem was further investigated in a number of publications (see [12], [13] and referencies therein). Here we put forward the results of [13].

Choose an orthogonal coordinate system $(x_1, x_2, x_3)$ such that $\boldsymbol{B}=B_0\boldsymbol{e_3}$ is directed along the $x_3$-axis and the wave vector $\boldsymbol{k}$ lies in the $(x_1, x_3)$-plane: $\boldsymbol{k}=(k_1, 0, k_3)$. The Hamiltonian function of the particle is

$$H = \sqrt{m^2c^4 + c^2p_1^2 + (cP_2 - eB_0x_1)^2 + c^2p_3^2} +$$
$$e\Phi_0 \cos(k_1x_1 + k_3x_3 - \omega t), \qquad (16)$$

where $P_2 = p_2 + eB_0x_1/c$, $\boldsymbol{p}=(p_1, p_2, p_3)$ is the particle's momentum, $m$ and $e$ are the particle's mass and charge accordingly. Introduce notations:

$$\omega_c = \frac{eB_0}{mc}, \quad k = (k_1^2 + k_3^2)^{1/2}, \quad \varepsilon = \frac{e\Phi_0}{mc^2},$$

$$\Omega_c = \omega_c/\varepsilon, \quad \sin\alpha = k_3/k.$$

We consider the case when the particle is relativistic and the wave is low-amplitude and high-frequency: $|\boldsymbol{p}|/(mc) \sim 1$, $\omega/(kc) \sim 1$, $\varepsilon \ll 1$, $\omega/\omega_c \sim \varepsilon$.

In the absence of the wave, the particle moves along a spiral: it rotates along the Larmor circle, and drifts along the magnetic field. The wave is high-frequency, and one can average over the wave's phase. Thus we obtain adiabatic motion, coinciding with the Larmor motion. However, in the process of slow Larmor motion the projection of the particle's velocity onto the direction of $\boldsymbol{k}$ is varying, and at certain time moments it can be equal to the wave's phase velocity. Near this resonance the argument of cosine in (16) is varying slowly, and the averaging does not work properly.

Our system with Hamiltonian (16) has 3½ degrees of freedom, but in the chosen system of coordinates the Hamiltonian does not depend on coordinate $x_2$, hence canonically conjugated momentum $P_2$ is constant, and we can put $P_2=0$. Now, if we introduce the wave's phase as a new variable, we get rid of explicit time-dependence and obtain a 2 d.o.f Hamiltonian system.

Rescale the variables:
$$\tilde{p}_{1,3} = p_{1,3}/(mc), \quad \tilde{q}_{1,3} = q_{1,3}/c, \quad \tilde{k}_{1,3} = k_{1,3}c,$$
$$\tilde{H} = H/(mc^2).$$

Now make a canonical transformation of variables $(q_1, p_1, q_3, p_3) \rightarrow (q, p, \varphi, I)$ (tildes are omitted) defined by generating function $W(q_1, q_3, I, p, t) = (k_1q_1 + k_3q_3 - \omega t)I + q_1 p$. Introduce notation $\hat{q} = \varepsilon q$. Omitting the hat, we obtain Hamiltonian function of the form

$$H = \sqrt{1+(kI+p\cos\alpha)^2 + p^2\sin^2\alpha + \Omega_c^2 q^2} - \omega I + \varepsilon\cos\varphi \equiv H_0 + \varepsilon\cos\varphi, \quad (17)$$

where conjugated variables are $(p, \varepsilon^{-1}q)$ and $(I, \varphi)$. This is a slow-fast system of the kind described in the previous sections, and even simpler, as we have here only one angle variable, and the resonance condition is $\dot{\varphi} = 0$. While $\dot{\varphi} \neq 0$, the particle's phase trajectory in the space $(p, q, I)$ lies in a vicinity of the intersection of a second order surface $H_0 = $ const and a plane $I = $ const. This intersection is an ellipse corresponding to the Larmor motion and represents an adiabatic trajectory. Resonant condition $\dot{\varphi} = \partial H_0/\partial I = 0$ defines in the $(p, q, I)$-space resonant surface $I = I_{\text{res}}(p, q)$. This surface turns out to be a sheet of a two-sheeted hyperboloid. The phase trajectory of adiabatic motion periodically crosses this surface. Trajectories of the resonant flow on the resonant surface are given by intersections of level surfaces $H_0 = $ const and the resonant surface. Hamiltonian of the resonant flow is

$$\Lambda = (\omega p/k)\cos\alpha + (1-(\omega/k)^2)^{1/2} \times \sqrt{1+p^2\sin^2\alpha+\Omega_c^2 q^2} \quad (18)$$

Trajectories of the resonant flow turn out to be planar sections of the resonant surface. If $\omega/(kc)<\sin\alpha$, they are ellipses, and if $\omega/(kc)>\sin\alpha$, they are hyperbolas. [For physical clarity, we write these conditions in dimensional units]. In the first case even if a phase point is captured into the resonance, it cannot go to infinity. In the second case a captured phase point can go to infinity, and this corresponds to unlimited surfatron acceleration. Indeed, as it can be seen from (17), if value of $I$ grows to infinity, the particle's energy also grows to infinity and the particle's velocity tends to the speed of light. As it was discussed in the previous section, the possibility of the capture and the unlimited acceleration depend on behaviour along the resonant flow of the area $S(p, q)$ of the oscillatory domain on the phase portrait of the pendulum-like system. After the procedure described above, we obtain the pendulum-like system in this case:

$$F = \frac{1}{2}gP^2 + \cos\varphi + b\varphi, \quad (19)$$

where

$$b = \frac{\Omega_c^2 \cos\alpha}{\sqrt{k^2-\omega^2}} \cdot \frac{q}{\sqrt{1+p^2\sin^2\alpha+\Omega_c^2 q^2}},$$
$$g = \frac{k^2(1-(\omega/k)^2)^{3/2}}{\sqrt{1+p^2\sin^2\alpha+\Omega_c^2 q^2}}. \quad (20)$$

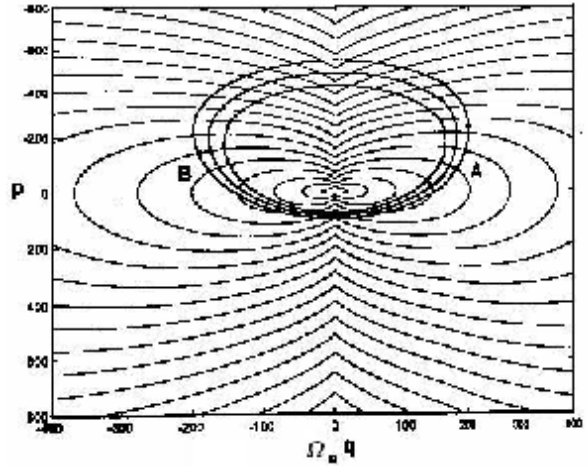

(a)

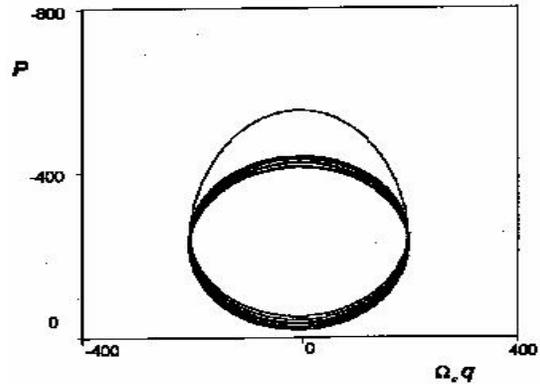

(b)

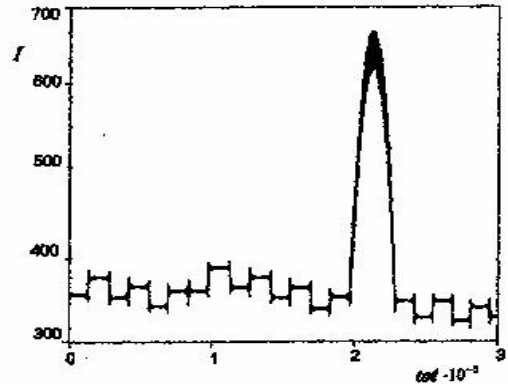

(c)

**FIGURE 2.** Motion with capture into the resonance and escape from the resonance, $\omega/(kc)<\sin\alpha$.

Using these formulae, one can study all possible cases of arrangement of level lines of $S(p, q)$ and trajectories of the resonant flow on the resonant surface (see [13]). Here we give two sample cases.

The resonant surface is uniquely projected onto $(p, q)$-plane, so we can study this projection. The first case $(\omega/(kc) < \sin \alpha)$ is shown in Fig. 2. Trajectories of the resonant flow are ellipses, shown in Fig.2 a with bold lines. Motion along these trajectories is directed counter-clockwise. The other family of curves in the figure represents level lines of area $S(p, q)$. Suppose a phase point hits the resonant surface at the point corresponding to $A$ in Fig.2 a. At this point the area $S(p, q)$ grows along the trajectory of the resonant flow, thus capture into the resonance is possible. If the point is captured, it moves along the ellipse on Fig.2 a. After a while, $S(p, q)$ stops growing and starts decreasing. At point $B$ the ellipse crosses again the level line of $S(p, q)$ containing $A$, and here the phase point escapes the resonance and leaves the resonant surface.

Projection of a phase trajectory onto $(p, q)$-plane in this case is shown in Fig.2 b. The phase point moves along the Larmor trajectory and repeatedly crosses the resonance. Probability of capture is small, but after a number of resonance crossings, the phase point is captured and moves along the resonant surface. Then it escapes from the resonance and continues its motion as before. Note, that at each resonance crossing value of adiabatic invariant $I$ undergoes a small jump, resulting from scattering on the resonances. The plot of variation of $I$ with time is presented in Fig.2 c.

In the second case, shown in Fig.3, trajectories of the resonant flow are hyperbolas. As in Fig.2, the other family of curves represents level lines of function $S(p, q)$. Under additional condition

$$\frac{E_0}{B_0} > \frac{\sqrt{(\omega/kc)^2 - \sin^2 \alpha}}{(\omega/kc)\sqrt{1 - (\omega/kc)^2}} \qquad (21)$$

these two families of lines are arranged like in Fig.3, and in the motion along a trajectory of the resonant flow area $S(p, q)$ always grows. Therefore, a captured phase point always stays captured and goes along one of the hyperbolas to infinity. Projection of a phase trajectory onto $(p, q)$-plane in this case is shown in Fig.3 b. The phase point rotates along a Larmor trajectory, repeatedly crossing the resonant surface, then it is captured and goes to infinity. Variation of $I$ with time is shown in Fig.3 c. At every crossing of the resonance the trajectory is scattered, and value of $I$ undergoes a small jump. When the trajectory is captured, $I$ tends to infinity. This corresponds to unlimited surfatron acceleration.

In Fig.4, variation of $I$ at a single scattering on the resonance is shown. Far from the resonance, the value of $I$ is aproximately conserved, oscillating a little. As the phase point approaches the resonance, amplitude of these oscillations grows, then a jump occurs, and after crossing the resonance, $I$ oscillates around a new value.

In conclusion we mention that the described method is quite general and can be applied to study resonant phenomena in slow-fast systems arising in various problems of physics. It allows for complete description of motion in the presence of isolated resonances. Certainly, one may encounter technical difficulties in analytical description of resonant flow and function $S(p, q)$, but they can be studied at least numerically. Other examples of systems where these methods can be applied include slowly perturbed billiards [14], slowly irregular waveguides [15], classical hydrogen molecular ion [16] and motion of charged particles in the Earth's magnetotail [17].

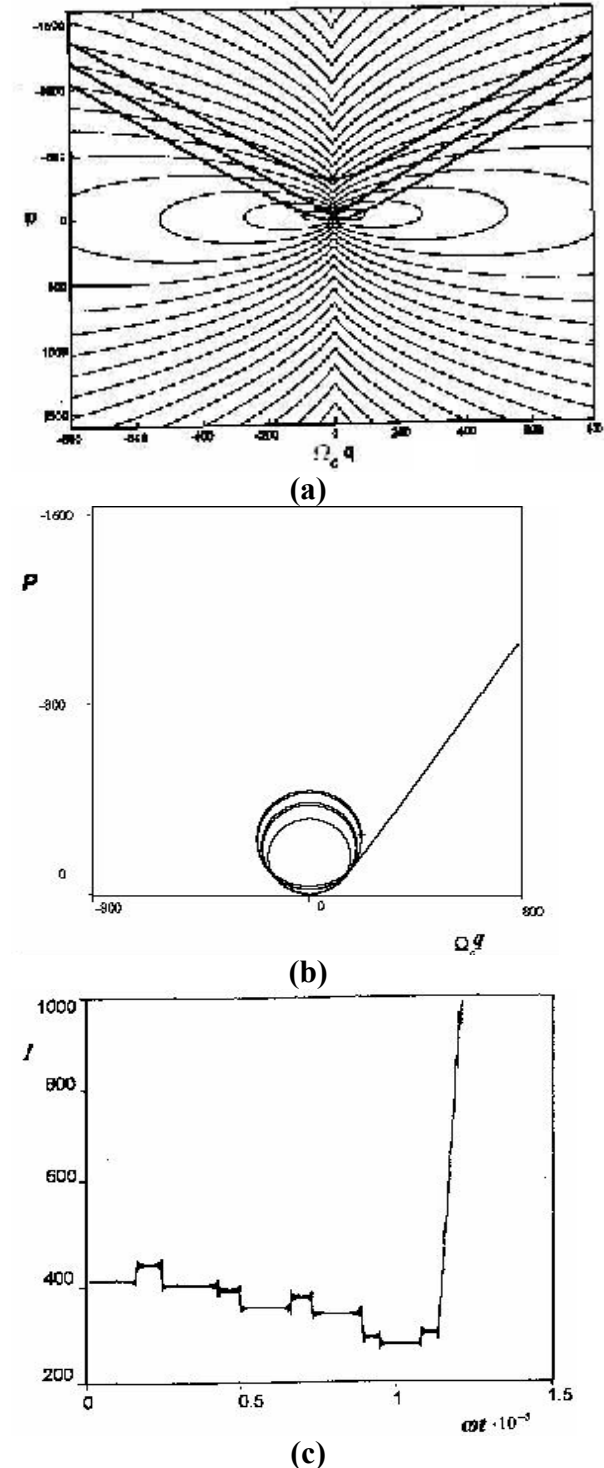

**FIGURE 3**. Motion with capture into resonance and unlimited acceleration, $\omega/(kc) > \sin \alpha$ and condition (21) is met.

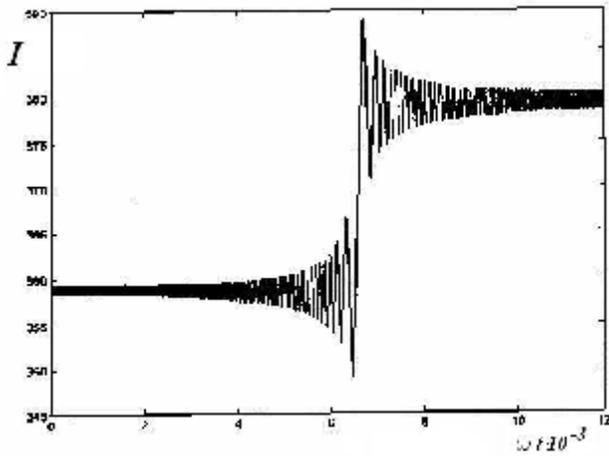

**FIGURE 4.** Variation of $I$ at scattering on the resonance.

## ACKNOWLEDGEMENTS

The work was paritally supported by RFBR grants 03-01-00158 and NSch-136.2003.1.